\documentclass[aps,prd,reprint,secnumarabic,nonbibnotes,floatfix,amssymb,longbibliography,showpacs,preprintnumbers]{revtex4-1}

\usepackage{graphicx}
\usepackage{subfigure}
\usepackage{mathptmx}
\usepackage{helvet}
\usepackage{lineno}
\usepackage{textcomp}
\usepackage{multirow}
\usepackage{amsmath}

\begin{document}

\title{ Study of the helix structure of QCD string}

\author{\v{S}\'{a}rka Todorova-Nov\'{a}}  
\affiliation{Tufts University, Medford MA, U.S.A.}
\email[sarka.todorova@cern.ch]{}

\begin{abstract}   
   The properties of a helix-like shaped QCD string are studied in the context
 of the Lund fragmentation model, where the concept of a 3-dimensional structure of the gluon field
 offers an alternative aproach to the modelling of the transverse momentum of hadrons.
 The paper is focused on the phenomenology of the model, which introduces correlations
 between transverse and longitudinal components of hadrons, as well as azimuthal ordering
 of hadrons along the string. The similarities between 2-particle correlations stemming
 from the helix-string structure and those commonly attributed to the Bose-Einstein interference are
 pointed out. It is shown that the charged assymetry observed in hadron production of close hadron pairs
 can be associated with fluctuations in the space-time history of the string breakup, and with the presence
 of resonant hadronic states.
 
\end{abstract}

\pacs{13.66.Bc,13.85.Hd,13.87.Fh}

\maketitle

\section{Introduction}

  The Lund string fragmentation model \cite{lund} 
 represents one of the most successful phenomenological descriptions 
 of the non-perturbative stage of the hadronisation
 process, where partons (quarks and gluons) are converted into hadrons.
 The understanding of hadronisation is essential for a reliable
 interpretation of the hadronic data obtained in collider experiments, and
 the PYTHIA generator code \cite{pythia} which contains the Monte-Carlo implementation
 of the Lund fragmentation model plays an important part in practically every analysis.
 
\begin{figure}[tbh]
\begin{center}
\includegraphics[width=0.45\textwidth]{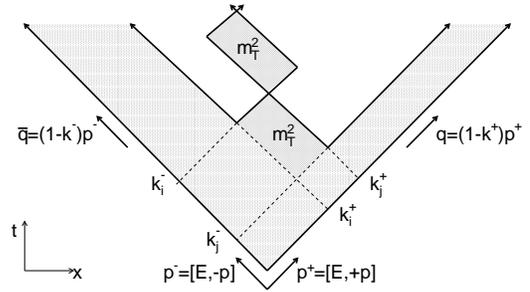}
\caption{ 
 Evolution of the QCD string in the rest frame
 of the $q\bar{q}$ pair. The endpoint partons loose their momentum as they
 separate and the string - the confining field - is created.
 The space-time coordinates of break-up vertices can be obtained from the relation
 [t,x]=(k$^+$p$^+$+k$^-$p$^-$)/$\kappa$ ($\kappa  \sim$ 1 GeV/fm), with $k^{+(-)}$ designating the fraction
 of parton momenta used for the build-up of the string field. 
 The x direction is parallel to the string axis and the mass of partons is neglected.
\label{fig:area}}
\end{center}
\end{figure}

    The model uses the concept of a string with uniform energy density to 
 model the confining colour field between partons carrying
 complementary colour charge. The string is viewed as being composed of straight
 pieces stretched between individual partons according to the colour flow.
 The fragmentation of the string proceeds via the tunneling effect (creation
 of a quark-antiquark pair from the vacuum) with a probability
 given by the fragmentation function. The space-time sequence of string break-up points defines the
 final set of hadrons, each built from a $q\bar{q}$ pair ( or a quark-diquark pair in the case
 of baryons) and a piece of string between the two adjacent string break-up points.
 The longitudinal hadron momenta stem directly from the space-time
 difference between these vertices, see Fig.~\ref{fig:area}.
  
  The idea of a helix-like shaped string was first suggested in the study of the properties
 of soft gluon emission  by Andersson et al.~\cite{lund_helixm}.
 Under the assumption that the generating current
 has a tendency to emit as many soft gluons as possible, and due to the  constraint 
 imposed on the emission angle
 by helicity conservation, it was shown that the optimal packing of emitted gluons
 in the phase space
 corresponds to a helix-like ordered gluon chain. Such a structure of the colour field cannot
 be expressed through gluonic excitations of the string and it needs to be implemented as an intrinsic string property.
  
  The current paper is organized as follows: section II describes the changes in the fragmentation model related to the helix string structure, the modification of the original helix string
 proposal, and the Monte-Carlo implementation of the model. Section III uses a simplified
 version of the hadronisation model for a detailed study of the experimental signature of
 the helix string model. Section IV deals with the charge asymmetry in the correlated
 hadron production. Section V provides an overview and discussion of the observed features. 
 Appendices A and B show the results of first tuning studies using the LEP and LHC data.

\section{Modelling of  transverse momentum}

\begin{figure}[hbt]
\begin{center}
\includegraphics[width=0.4\textwidth]{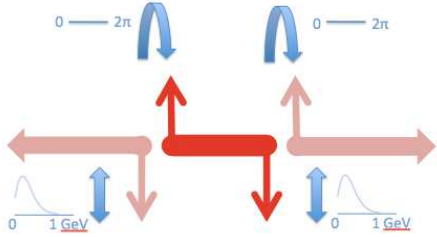}
\caption{ The generation of the transverse momentum of a hadron in the conventional Lund string model, obtained as a vectorial sum of the transverse momenta of quarks(diquarks) created in the tunneling process. The 
string breakup points are uncorrelated in the azimuthal angle and the (locally conserved) transverse momentum
is sampled from a gaussian distribution.
\label{fig:lund_pt}}
\end{center}
\end{figure}

   The implementation of a string with a helix structure radically
 changes the way hadrons acquire their transverse momentum.
 In the conventional Lund model \cite{lund}, the transverse momentum
 of the hadron is the (vectorial) sum of the transverse momenta of the
 (di)quarks which were created via tunneling during the
 breakup of the string. The transverse momenta of newly created partons
 are randomly sampled from a gaussian distribution (with adjustable width) and their
 azimuthal direction is uniformly random (i.e. two random numbers are thrown at 
 every string break-up point in order to generate the associated transverse momentum, Fig.~\ref{fig:lund_pt}).

    In the case of the helix ordered string, hadrons obtain their transverse momentum from the shape 
 of the colour field itself, so that there is in principle no need to assign a momentum 
 to new quarks in the string breakup.  If we assume the gluon field has a form of an ideal helix 
 and the string tension is tangential to the helix, the transverse
 momentum stored in the string piece defined by two adjacent string break-up points can be written as

\begin{equation}
    \vec{p}_{\rm T} = R \int_{\Phi_i}^{\Phi_j}\exp^{i(\Phi\pm\pi/2)}d\Phi 
\end{equation}

   where $\Phi_{i(j)}$ is the 'phase' of the helix (azimuthal angle)
 at the break-up point $i(j)$ and $R$ stands for the radius of the helix.

\begin{figure}[hbt]
\begin{center}
\includegraphics[width=0.3\textwidth]{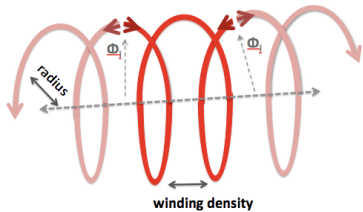}
\caption{ In the helix string model, the transverse momentum is stored in the transverse
 structure of the gluon field.  The tunneling process
 creates quark-antiquark pairs at rest and the string breakup points are correlated. 
\label{fig:helix_pt}}
\end{center}
\end{figure}

  The transverse momentum that the hadron carries is entirely defined by the properties of the helix
 field (described by two parameters: the radius  and the winding density, see Fig.~\ref{fig:helix_pt}). The loss of the azimuthal degree of freedom in the string break-up is arguably the most significant consequence of the inclusion of the helix string model
 in the fragmentation process.

\subsection{Parametrisations of the helix string structure}

\begin{figure}[b!]
\begin{center}
\includegraphics[width=0.4\textwidth]{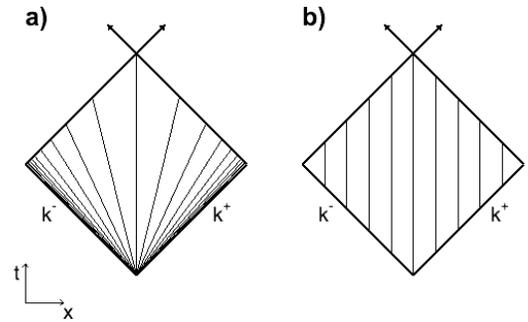}
\caption{
 The space-time evolution of the helix string structure in the rest frame of a $q\bar{q}$ string.
 Left: original proposal (Eq.\ref{eq:lund_helix}), where helix phase evolves along the string axis in time.
 Right: modified proposal (Eq.\ref{eq:my_helix}), corresponds to a static helix structure. 
\label{fig:phase}}
\end{center}
\end{figure}

   In the original helix string proposal \cite{lund_helixm}, the phase difference of the helix winding
 is related to the rapidity difference of the emitting current by the formula:
 
\begin{equation}
        \Delta\Phi = \frac{\Delta y}{\tau},
\label{eq:lund_helix}
\end{equation}

 where $\Delta\Phi$ is the difference in helix phase between two points along the string,
 $\tau$ is a parameter, and  $\Delta y$ is the rapidity difference which can be calculated as
\begin{equation}
        \Delta y = 0.5 \ {\rm ln}(\frac{k_i^+ k_j^-}{k_i^- k_j^+}),
\label{eq:dy}
\end{equation}
   with $k^{+,-}$ representing the fractions of endpoint quark momenta
 defining a position along the string, see Fig.\ref{fig:area}. In this parametrization,
 the helix phase difference is related to the {\it angular} difference of points 
 in the string diagram (Fig.\ref{fig:phase}a). 

   An alternative parametrisation \cite{modified_helix} sets 
 the difference in the helix phase to be proportional to the energy stored in between
 two points along the string
\begin{equation}
     \Delta \Phi = {\cal S} \ (\Delta k^+ + \Delta k^-) \ M_0/2 ,
\label{eq:my_helix}
\end{equation}
  where $M_0$ stands for the invariant mass of the string, $\cal S$[rad/GeV] is the density of the helix winding,
 and $\Delta k^+= |k^+_j - k^+_{j+1}|$, $\Delta k^-= |k^-_j-k^-_{j+1}|$ define the size of the string piece. This parametrisation describes a static helix structure where the helix phase
 remains constant at a given point along the string (Fig.\ref{fig:phase}b). As discussed in Section 3.2 of~\cite{modified_helix}, such a parametrisation emerges as a solution of equations developed in~\cite{lund_helixm} assuming the separation of soft gluons is dominated by the azimuthal component (neglected in~\cite{lund_helixm}).

 For practical reasons, only the static
 helix scenario (Eq.~\ref{eq:my_helix}) is studied in this paper - the original helix string proposal
 cannot be implemented for an arbitrary parton configuration without modifications 
 ( the model contains a singularity at the gluon kink, where one of the fractions $k$
  becomes 0 in Eq.~\ref{eq:lund_helix} ).

  For simplicity, the static helix scenario will be refered to as ``helix string model'' in the following,
 even though it represents just one possible solution for the helix string parametrisation.
 The phenomenology of the helix string depends on the exact definition of its shape and the experimental
 signatures may vary to some extent.

\subsection{ Model implementation }

  The helix string fragmentation model has been implemented in a PYTHIA-compatible form, and
 the fragmentation code of PYTHIA \cite{pythia8} has been adapted \cite{pystrf} to allow
 switching between the standard Lund string fragmentation and 
 the helix string fragmentation (refered to as ``standard'' and ``helix'' fragmentation, respectively).
   
\begin{figure}[b!]
\begin{center}
\includegraphics[width=0.3\textwidth]{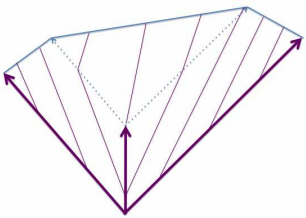}
\caption{   An illustration of the helix phase evolution of the modified helix model in the presence of a hard gluon kink on the string.
\label{fig:kink}}
\end{center}
\end{figure}

  The comparison of the helix string model with data requires the model to be extended to cover
 not only the simple case of a $q\bar{q}$ system but also an arbitrary multiparton configuration corresponding to the emission
 of hard gluons from the quark-antiquark dipole, as in Fig.~\ref{fig:kink}. This is actually the most complicated part of the model implementation which requires some additional assumptions to be made. 
    
  First, it is assumed that the helix phase
 runs smoothly over the gluon kink, i.e. the helix phases at the connecting ends of adjacent string pieces coincide.

  Second, it is assumed that the probability of the soft gluon emission depends on the mass of the string piece
 formed by colour partners in the quark-gluon cascade 
 (the gluon momentum is equally split between the two adjacent string pieces), so that
 the helix phase difference between string endpoints, for a colour ordered system of
 N partons, becomes
\begin{equation}
  \Delta\Phi = {\cal S} \sum_{i}^{i<N} M_{i},
\label{eq:combphase}   
\end{equation} 
  where the sum runs over all (ordered) string pieces and $M_i$ is the mass of the i-th string piece.
 In the simulation, the random choice of the helix phase at one point along the string defines
 the helix phase for the entire space-time history of the string evolution. In particular, 
 the knowledge of the combined phase difference allows the fragmentation to proceed in the
 usual way, by steps taken randomly from one of the string's ends.

\section{Experimental signature}

 For a better understanding of the origin of observable effects, it is useful to consider a simplified model
 configuration  with ideal helix shape which provides a better illustration of the basic features of the model.   
 The smearing of the helix shape can be obtained from the comparison with the data (Appendix A).

\begin{figure}[h!]
\begin{center}
\includegraphics[width=0.4\textwidth]{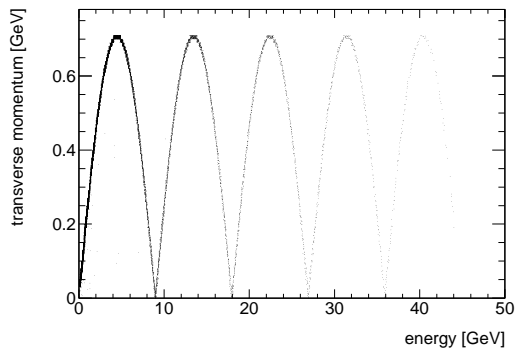}
\caption{ The dependence of the direct hadrons transverse momentum on their
 energy in the rest frame of the fragmenting helix string with radius $R$=0.36 GeV and
 $\cal{S}$=0.7 rad/GeV. Toy model of hadronic $Z^0$ decay with suppressed parton showering.
\label{fig:pt_e_direct}}
\end{center}
\end{figure}

  The correlations between transverse and longitudinal components of direct hadron momentum which
 originate from the helix string structure are best visible in a toy study of fragmentation
 of a simple $q\bar{q}$ string, shown in Fig.~\ref{fig:pt_e_direct}. The longitudinal direction
 and the transverse plane are defined with respect to the string axis.
 
\begin{figure}[b!]
\begin{center}
\includegraphics[width=0.4\textwidth]{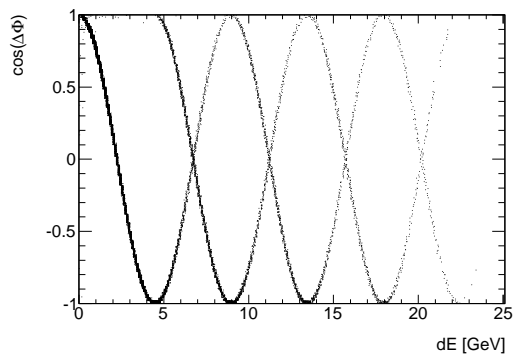}
\caption{ The dependence of the azimuthal opening angle of two direct hadrons on their energy-distance $dE$
 along the string in the rest frame of the fragmenting helix string with parameter $\cal{S}$=0.7 rad/GeV.
 Toy model of hadronic $Z^0$ decay with suppressed parton showering. For combination of hadrons 
 with energy distance exceeding a half-loop of the helix winding, two possible configurations exist.  
\label{fig:dphi_de_direct}}
\end{center}
\end{figure}

  The helix string structure also creates a very distinct correlation between the azimuthal opening angle of two direct hadrons and their distance along the string, which can be expressed as a function of the amount of energy stored in the ordered hadron chain separating the pair, see Fig.~\ref{fig:dphi_de_direct}. The position of the hadron is associated with the middle point of the string piece which forms the hadron.

  Both types of correlations are rather difficult to observe experimentally, though.
 In the presence of gluon kinks, it is impossible to define the longitudinal
 axis of the whole string in a way which allows the separation of the intrinsic longitudinal and transverse
 momenta of all hadrons.  On top of this important source
 of smearing, it is necessary to account for the presence of non-direct hadrons (resonance decay products),
 or even - as in the case of hadron-hadron collisions - several overlapping hadron systems, which contribute
 considerably to the dilution of the features associated with the underlying helix string structure. 
    
  At this point, it is instructive to separate the effect of the parton showering (and string overlap )
 from the effect of resonance decay. The next step therefore consists in
 the study of a realistic configuration of input partons (quarks and hard gluons), but with fragmentation
 limited to the production of direct charged pions ($\pi^+$, $\pi^-$). 

\begin{figure}[h]
\begin{center}
\includegraphics[width=0.4\textwidth]{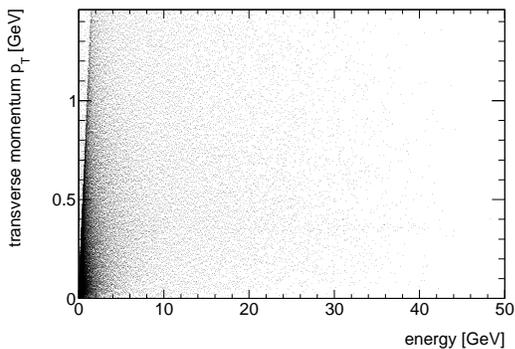}
\caption{ The dependence of the size of the transverse momentum(*) of the direct pions on their
 energy in the $Z^0$ rest frame. Hadronic $Z^0$ decay with suppressed resonance production.
 The underlying correlations are smeared by the parton shower.(*) transverse plane is defined with respect to the
 event Thrust axis. 
\label{fig:pt_e_shower}}
\end{center}
\end{figure}

  Figure~\ref{fig:pt_e_shower} shows the disappearance of the clear interference pattern between the transverse
 momentum and the energy of a direct hadron in a sample of hadronic $Z^0$ decays into light quark pairs ($u\bar{u}$, $d\bar{d}$), with parton shower enabled and resonance production disabled. The creation of heavier quarks in the parton shower and in the tunneling is suppressed, too.
 The difference between Figs.~\ref{fig:pt_e_direct}
and~\ref{fig:pt_e_shower} is entirely attributed to the presence of a hard gluon cascade and the associated
 complex string topology. 

\begin{figure}[t!]
\begin{center}
\includegraphics[width=0.4\textwidth]{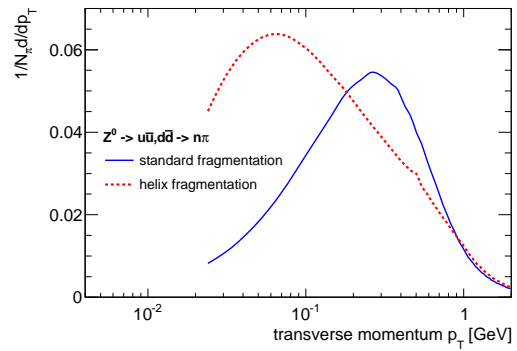}
\caption{ The comparison of the inclusive transverse momentum (*) distribution obtained via standard
 fragmentation (full blue line) and via helix string fragmentation (dashed red line).
 Hadronic $Z^0$ decay with suppressed resonance production.(*) transverse plane is defined with respect to the
 event Thrust axis. 
\label{fig:pt_inclusive}}
\end{center}
\end{figure}
 
  Even though the correlation pattern is smeared out, the presence of underlying correlations is 
 still visible - indirectly - in the shape of the inclusive p$_{T}$ spectrum.
 Figure~\ref{fig:pt_inclusive} shows the comparison of normalized p$_{T}$ spectra for the
 standard fragmentation and the helix string fragmentation, in the same sample of Z$^0$s decaying, ultimately,
 into a set of direct pions. The radius of the helix string has been adjusted ($R=0.5$ GeV) in order to reproduce
 the average $<p_T>$ obtained in the conventional fragmentation (with $\sigma_{p_T}=0.36$ GeV).  The parameter $\cal{S}$ is set to 0.7 rad/GeV and kept unchanged in this paper.
 The difference in the shape of the two distributions reflects the difference in model
 assumptions - the standard Lund fragmentation creates a gaussian-like shape while in the helix string fragmentation,
 the shape of the inclusive p$_T$ is closer to an exponentially falling function. The study of the shape
 of the inclusive $p_T$ spectrum is one possible way how to experimentally distinguish between the two model variants. 
 Note that the difference between models is artificially enhanced in the sample where all hadrons are direct.

\begin{figure}[b!]
\begin{center}
\includegraphics[width=0.4\textwidth]{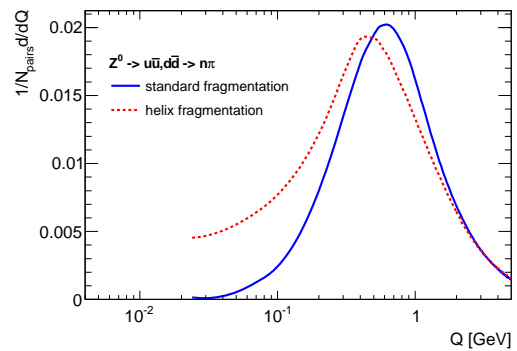}
\caption{ The comparison of the momentum difference distribution for pairs of charged pions obtained via standard
 fragmentation (full blue line) and via helix string fragmentation (dashed red line).
 Hadronic $Z^0$ decay with suppressed resonance production.
\label{fig:q_inclusive}}
\end{center}
\end{figure}

  A similar pattern is found in the study of 2-particle correlations, where the helix string structure
  causes an enhanced production of pairs of close direct hadrons. This is seen in Fig.~\ref{fig:q_inclusive} 
  in the distribution of the momentum difference Q defined as

\begin{equation}
   Q = \sqrt{ - ( p_1 - p_2 )^2 },
\end{equation}  
  where p$_i$ stands for the 4-momenta of hadrons. The comparison is done using
 pairs of charged hadrons ( direct pions in our example ).

  The traces of the 2-particle correlation pattern shown in Fig.~\ref{fig:dphi_de_direct} can
 also be detected with the help of a suitably defined power spectrum on a selection of events with
 restricted parton shower activity. Indeed, this is what has been found in a recent study by ATLAS \cite{ao},
 where the sensitivity to the parameters of the
 helix string structure has been investigated on MC in low p$_{T}$ inelastic proton-proton scattering.

\section {Charge asymmetry} 

\begin{figure}[bth]
\begin{center}
\includegraphics[width=0.49\textwidth]{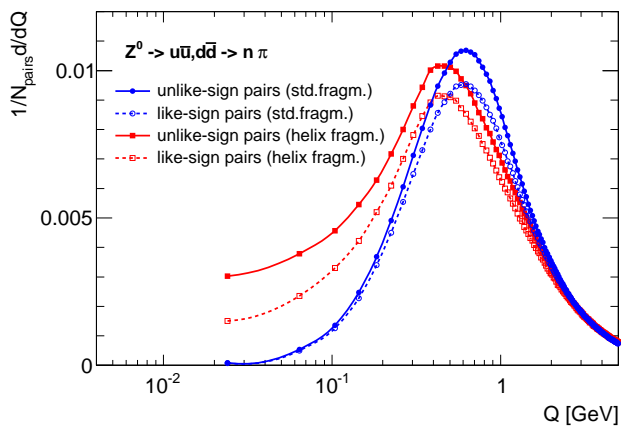}
\caption{ The comparison of the momentum difference distribution for pairs of charged pions obtained via standard
 fragmentation (blue lines) and via helix string fragmentation (red lines). The contributions from pairs with identical
 charge (``like-sign'') are indicated by dashed lines, the unlike-sign charge combinations correspond to full lines.
 Hadronic $Z^0$ decay with suppressed resonance production. 
\label{fig:q_charged}}
\end{center}
\end{figure}

  The prediction of low Q enhancement by the helix string
 model becomes even more intriguing when looking at the difference
 between pairs of hadrons with like-sign and unlike-sign charge combination - the enhancement
 is significantly more pronounced for the unlike-sign pairs
  (Fig.~\ref{fig:q_charged}). 

   A charge-asymmetric enhancement of 2-particle
 production at low Q is readily observed in hadronic data. However, this enhancement
 is seen in \emph{like-sign} pairs and usually associated
 with the Bose-Einstein interference, though the phenomenon has never been successfully simulated from first principles, and a number of measurements actually contradict this interpretation 
 (absence of correlations between overlapping hadronic systems~\cite{be_ww_ALEPH},~\cite{be_ww_DELPHI},~\cite{be_ww_L3},~\cite{be_ww_OPAL},
 dependence of the size of the ``source'' on the particle type ~\cite{be_fd_review}).

   The presence of string-structure induced 2-particle correlations    
  at low Q therefore raises questions : can these two observations
 be connected ? is there a way to modify the charge flow in the model
 so that it reproduces the charge asymmetry in the data ?
 
  It turns out there are actually two ways how to influence the charge asymmetry
 of 2-particle correlations  in the fragmentation of the helix string, and both represent
 a natural extension of the helix string model, as explained below.

 \subsection{Fluctuations in the space-time sequence of string breakup}

\begin{figure}[h!]
\begin{center}
\includegraphics[width=0.49\textwidth]{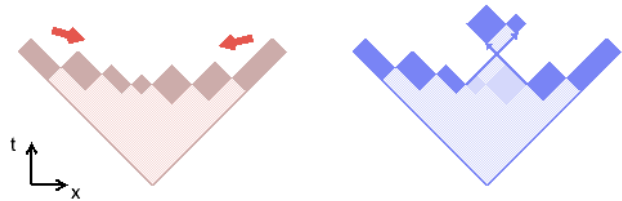}
\caption{ The string area diagram showing the PYTHIA fragmentation of $q\bar{q}$ string
 in an outside-in direction, creating a single hadron at each step (left diagram). The algorithm
 neglects fluctuations in the space-time sequence of the string breakup (right diagram) because 
 for the standard Lund string model, the final states resulting from both diagrams
 are indistinguishable in the energy-momentum space. The horizontal axis shows
 the distance along the string axis, the vertical axis indicates the time in the
 rest frame of the string. 
\label{fig:fluctuations}}
\end{center}
\end{figure}

  The PYTHIA generator fragments a QCD string from both of its endpoints, creating one direct
 hadron at each step, until the mass of the remaining piece of string falls below a certain
 (adjustable) limit, and is then split into two final hadrons. This is a convenient way of
 implementing the fragmentation algorithm that 
 neglects fluctuations in the lifetime of a string piece comprising several
 adjacent direct hadrons, as shown in Fig.~\ref{fig:fluctuations}. 

\begin{figure}[ht!]
\begin{center}
\includegraphics[width=0.4\textwidth]{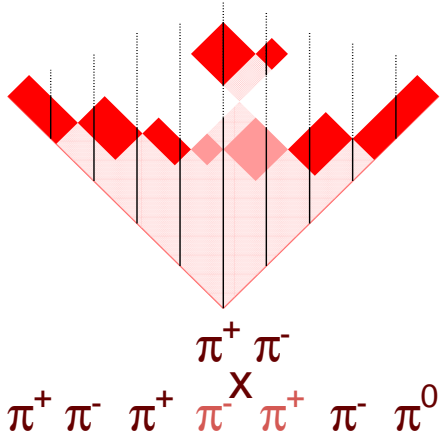}
\caption{ The string area diagram showing the effective swap of adjacent hadrons in the fragmentation of a QCD string
 with helical structure ( the vertical lines in the diagram show the evolution of a fixed helix phase).
 Since the transverse momentum of hadrons is associated with the transverse structure of the string,  
 the fluctuations of the space-time sequence of string breakup are reflected in the relative ordering of final hadrons in the azimuthal
 angle, creating distinguishable hadronic final states.
\label{fig:fluct_helix}}
\end{center}
\end{figure}

 It is argued in \cite{lund},section 2, that neglecting these fluctuations does not affect the outcome of the
 modelling since the hadron final states are identical in the energy-momentum space.
 This argumentat is indeed valid for the standard Lund string model, where the
 transverse momentum of direct hadrons is acquired in the tunneling process and
 carried by constituent quarks (or diquarks). However, if we consider the helix string
 model, where the transverse momentum is carried by the string itself, the fluctuations
 do have an impact on the final hadronic system -
 the hadrons become effectively reordered along the helix gluon chain. This is illustrated
 in Fig.~\ref{fig:fluct_helix}.

\begin{figure}[bth]
\begin{center}
\includegraphics[width=0.4\textwidth]{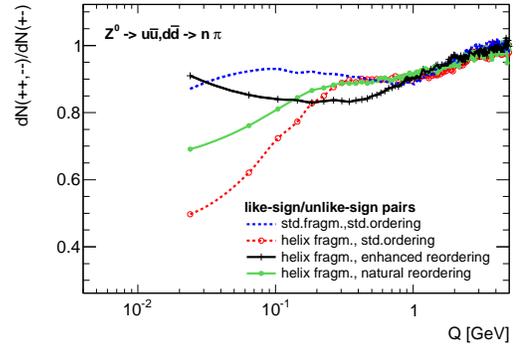}
\caption{ The relative fraction of like-sign and unlike-sign pairs of charged pions as a function of their momentum
 difference Q, of the string type and of the ordering of hadrons along the string. The helix-string fragmentation
 favours the creation of close adjacent unlike-sign hadron pairs in the conventional PYTHIA ordering (dashed lines).
 The asymmetry is attenuated in the presence of fluctuations which correspond to an effective reordering of hadrons
 along the gluon chain (full lines). Hadronic $Z^0$ decay with suppressed resonance production. 
\label{fig:q_charged_helix}}
\end{center}
\end{figure}

  The impact of these fluctuations ( i.e. the presence of the string breakup
 configurations which are omitted in the standard PYTHIA fragmentation chain ) on the relative
 contribution from like-sign and unlike-sign pairs of hadrons
 is shown in Fig.~\ref{fig:q_charged_helix}. The ratio of like-sign and unlike-sign Q distributions is plotted.
 Two scenarios are studied: the ``enhanced'' one, where the ordering is done in a way
 which favours the creation of ``adjacent'' like-sign pairs of hadrons, and the ``natural'' one,
 where the fluctuations occur randomly according to the string area law \cite{lund}. 
 In both scenarios, only the fluctuations relevant to the swap of adjacent (next-in-rank) hadrons
 are considered, and for the toy model of Z0 decay into direct pions, the average probability
 of the swap is around 0.4. For comparison, the ratios obtained with the standard PYTHIA ordering are
 included in the plot both for the standard Lund string and the helix string.

\subsection{Resonances}  

  The presence of resonant hadronic states in the fragmentation chain has a significant impact on
 the size of correlations stemming from the underlying string structure and on the charge asymmetry.
 The principal question resides in the treatment of the resonance decay - the resonance, in particular
 a short-lived one, can be viewed as a piece of the string which preserves its properties, and decays
 according to string break-up rules, as a smooth continuation of the string fragmentation process. It is also
 possible that resonances, especially long-lived ones,  gradually ``lose'' the memory of the QCD field.

\begin{figure}[bth]
\begin{center}
\includegraphics[width=0.4\textwidth]{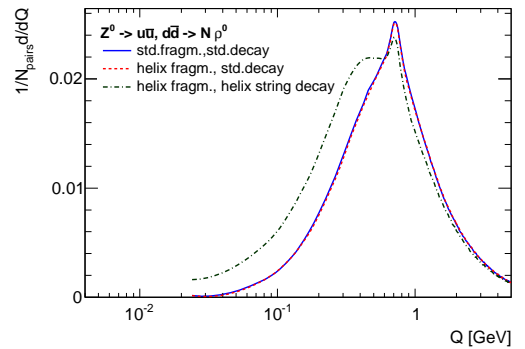}
\caption{ The comparison of the momentum difference distribution for pairs of charged pions obtained via standard
 fragmentation and standard resonance decay (full blue line), via helix string fragmentation and standard resonance decay
(dashed red line), and the helix string fragmentation incorporating the resonance decay (dotted-dashed green line).
 Hadronic $Z^0$ decay with forced $\rho^0$ resonance production. 
\label{fig:q_rho}}
\end{center}
\end{figure}

   To see the impact of the different models of the resonance decay on 2-particle correlations associated
 with the helix structure of the string, another version of the toy modelling of hadronic Z$^0$ decay
 is set up, where the QCD strings are forced to decay into a set of direct $\rho^0$ resonances. The $\rho^0$s
 consequently decay according to the usual PYTHIA decay algorithm (``std. $\rho^0$ decay''), or their decay
 obeys the helix string break-up rules (``helix $\rho^0$ decay''). The comparison of the momentum difference spectra
 for pairs of charged pions obtained with standard PYTHIA fragmentation and resonance decay, and with helix string
 fragmentation scenario combined with standard, resp.~helix resonance decay, is shown in Fig.~\ref{fig:q_rho}.
  It is not too surprising to see that the hadronisation scenario where the resonance decays do not take into account
 the underlying string structure does not exhibit the sort of enhancement observed with direct pions - the final hadrons
 ``lost'' the memory of the generating QCD field in the decay of the intermediate resonance state. It is also clear
 that the presence of resonances has a dampening effect on the low Q enhancement: the offspring of a resonance decay
 has a momentum difference $Q = \sqrt{M_{\rho_{0}}^2-4 m_{\pi}^4}$ so that the unlike-sign Q distribution contains
 a resonance structure, and the presence of a heavy resonance increases the average distance -along the string- between
 other direct hadrons (compare Fig.~\ref{fig:q_rho} and Fig.~\ref{fig:q_inclusive}).              

\begin{figure}[t]
\begin{center}
\includegraphics[width=0.4\textwidth]{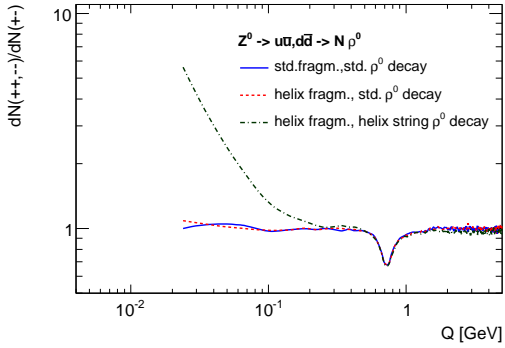}
\caption{ The ratio of the momentum difference of like-sign and unlike-sign pairs of charged pions obtained via standard
 fragmentation and standard resonance decay (full blue line), via helix string fragmentation and standard resonance decay
(dashed red line), and the helix string fragmentation incorporating the resonance decay (dotted-dashed green line).
 Hadronic $Z^0$ decay with forced $\rho^0$ resonance production. 
\label{fig:q_rho_ratio}}
\end{center}
\end{figure}
 
\begin{figure}[b!]
\begin{center}
\includegraphics[width=0.4\textwidth]{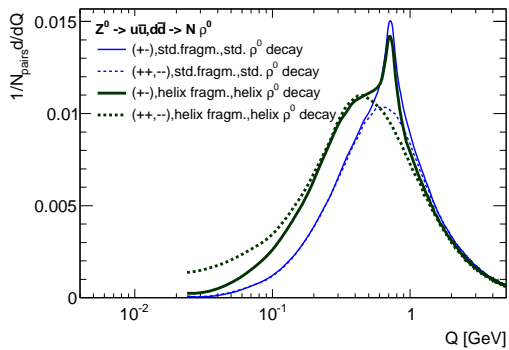}
\caption{ The comparison of the momentum difference distribution for pairs of charged pions obtained via standard
 fragmentation and standard resonance decay (blue lines) and the helix string fragmentation incorporating the resonance decay (green lines). The full lines indicate contributions from unlike-sign pairs, the dotted lines correspond to the contributions from like-sign pairs
of pions. Hadronic $Z^0$ decay with forced $\rho^0$ resonance production. 
\label{fig:q_rho_charged}}
\end{center}
\end{figure}

  It is however quite interesting to see that the inclusion of resonances enhances the relative fraction of
 like-sign and unlike-sign pion pairs at low Q, as shown in Fig.~\ref{fig:q_rho_ratio}.
 The effect is substantial and provides
 answers to questions formulated in the current section - the helix string model predicts a low Q
 enhancement dominated by like-sign pairs in the presence of resonances which preserve the ``memory'' of the gluon field
 through their decay.

  Figure~\ref{fig:q_rho_charged} compares the Q spectra, to stress the fact that even if 
 the like-sign pair production dominates the low Q region, some enhancement is also expected for the unlike-sign pairs
 when comparing the helix string scenario with the standard Lund fragmentation. This feature may become instrumental
 in the test of the model against the data. 

\begin{figure}[b!]
\begin{center}
\includegraphics[width=0.4\textwidth]{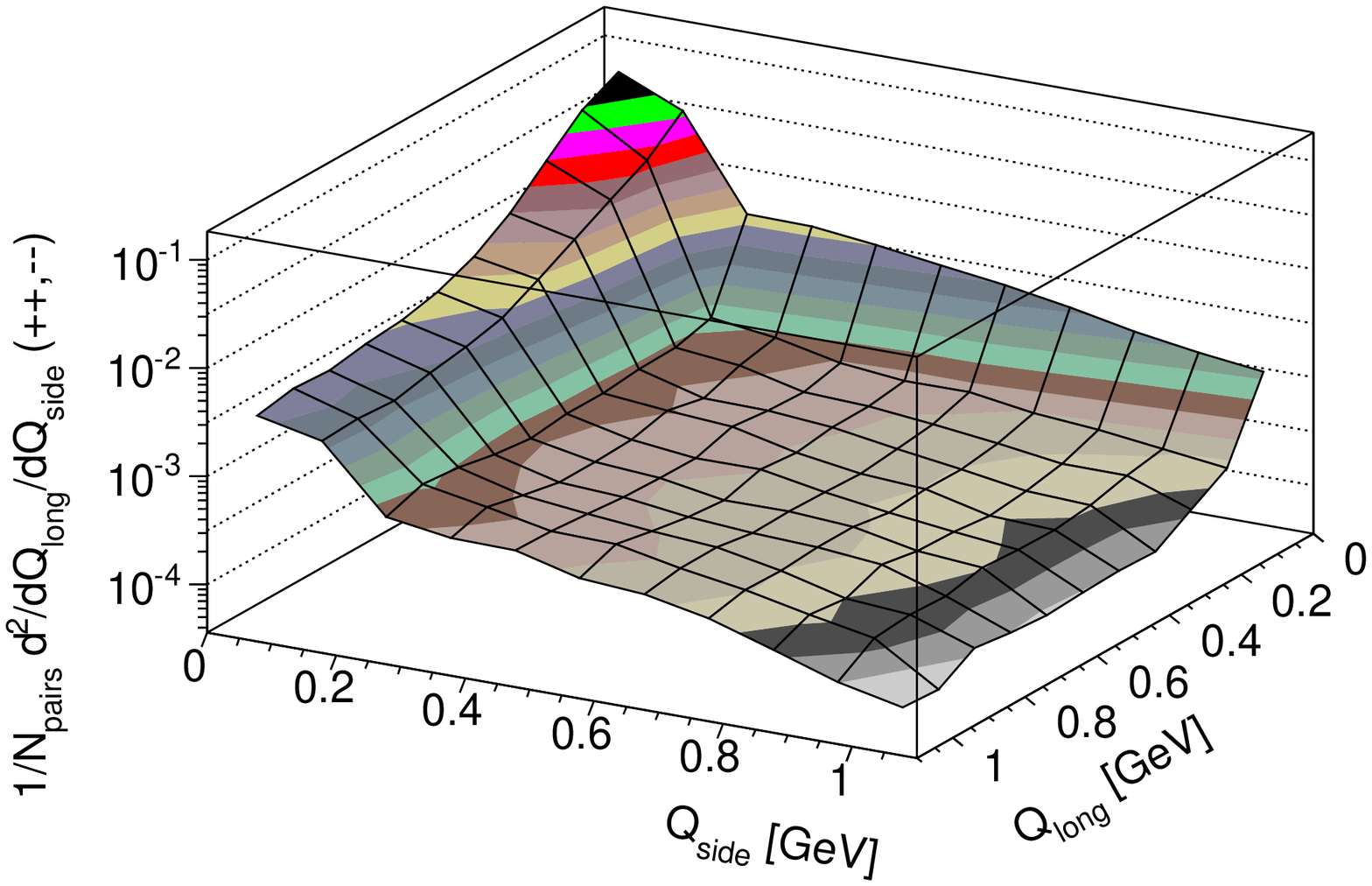}
\includegraphics[width=0.4\textwidth]{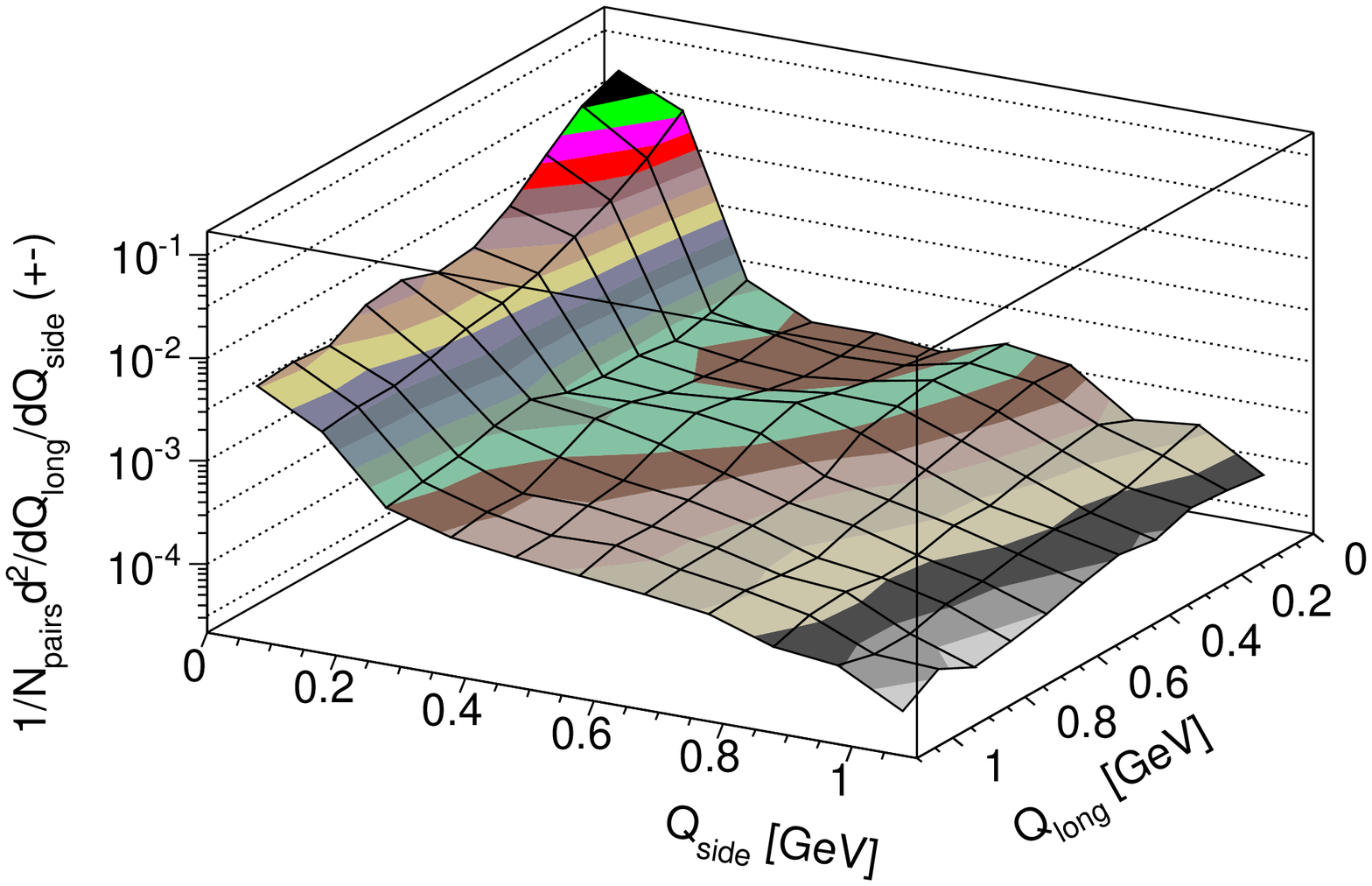}
\caption{ The comparison of the 2-dimensional momentum difference of like-sign (top) and unlike-sign (bottom) pion pairs obtained with the helix string fragmentation model.
 $Q_{long}$ ($Q_{side}$) correspond to the longitudinal(transverse) component of the momentum difference. 
 The circular band in the bottom spectrum is associated with the resonance decay. A clear enhancement of the $Q_{side}$ arm
 is visible in the like-sign spectrum. 
 Hadronic $Z^0$ decay with disabled parton shower and forced $\rho^0$ resonance production. 
\label{fig:elong_input}}
\end{center}
\end{figure}

\subsection{Elongated source?}

  Having successfully demonstrated - with the help of simplified fragmentation scenarios - that the enhanced production of like-sign
 hadron pairs with a low momentum difference is compatible with the helix string model, the temptation is great to take a step further
 and to try to see if the model predicts some of non-trivial properties of these correlations. In the analysis of
 the LEP data~\cite{l3},
 it has been shown that the correlation function exhibits an asymmetry in the longitudinal and transverse direction - the
 phenomenon is known as 'elongation' of the source in the terminology of Bose-Einstein interferometry. The asymmetry is seen as
 faster dampening of the correlation function along the longitudinal axis (the Thrust axis of the event).

\begin{figure}[h!]
\begin{center}
\includegraphics[width=0.49\textwidth]{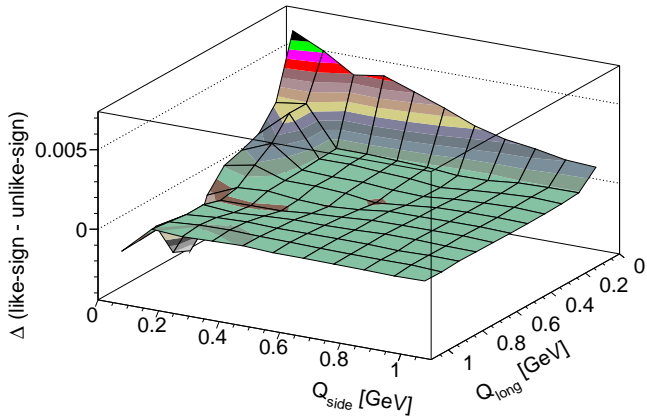}
\caption{ The difference of the 2-dimensional Q distributions of like-sign and unlike-sign pion pairs
 obtained with the helix string fragmentation model
 exhibits an asymmetry which may be at origin of so-called ``elongation'' effect.
 Hadronic $Z^0$ decay with disabled parton shower and forced $\rho^0$ resonance production. 
\label{fig:elongation}}
\end{center}
\end{figure}

   Once more, the toy model of hadronic Z$^0$ decay with forced $\rho^0$ production and subsequent helix-like decay
 of $\rho^0$ resonance into charged pions (the model which exhibits significant enhancement of like-sign pair production at low Q) is used,
 but the parton shower is switched off to prevent the smearing of transverse and longitudinal components. The difference between
 charged pion momenta is calculated in the longitudinal centre-of-mass frame of the particle pair (see~\cite{l3} for definition).
 Figure~\ref{fig:elong_input} shows the 2-dimensional distributions of like-sign and unlike-sign combinations. The circular band which is
 visible in the unlike-sign spectrum corresponds to the $\rho$ resonance decay. The like-sign spectrum exhibits an enhanced particle
 production in low $Q_{long}$ region which is not visible in the unlike-sign spectrum, and which may be at the origin of the elongation 
 effect (the detailed discussion of this feature is left to a separate publication). Indeed, after subtraction of the two distributions,
 the resulting picture shows a certain 'elongation' of the particle production, in any case an asymmetry in comparison of the shape
 of the like-sign pair excess in transverse and longitudinal directions (Fig.~\ref{fig:elongation}).   

\section{Overview}

   There are only a couple of free parameters in the helix string model: the helix radius $R$ (which
 replaces the $\sigma_{p_T}$ parameter in PYTHIA) and the density of the helix winding (parameter
 $\cal{S}$ in the case of the static helix scenario). The predictive power of the model has so far
 been used essentially for the development of the model itself : the necessity to accommodate
 the charged asymmetry in the predicted 2-particle correlations leads to the reconsideration of
 ``hidden'' degrees of freedom of the string fragmentation model (the dependence on the space-time
 history of the string breakup). The existence of these degrees of freedom translates into
 uncertainty of the model prediction which hopefully will be reduced by comparison with
 the experimental data.

   The key ingredient of the successful implementation of the helix model seems to be
 the adjustment of the relative fraction of direct hadrons, in particular short-lived
 resonances with decay conforming to the helix string fragmentation rules, and their ordering
 along the string. The suitable algorithm has yet to be developed, but some estimates
 of the observable effects can be made using a mixture of ``ordinary'' hadron production,
 and enhanced resonance production.

\begin{figure}[hbt]
 \begin{center}
 \includegraphics[width=0.4\textwidth]{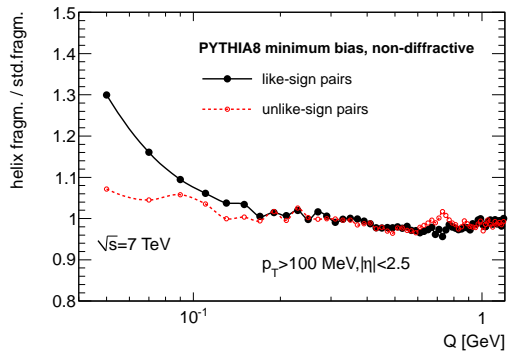}
 \caption{ An estimate of the contribution of the helix-string fragmentation model to the 2-particle correlation function
 in inelastic low p$_T$ pp collisions at LHC, using PYTHIA8 generator with the helix-string option \cite{pystrf}.
 \label{fig:mb_corr}}
 \end{center}
 \end{figure}

   An example of such an estimate is given in Fig.~\ref{fig:mb_corr}, for the study of the 2-particle correlations
 at LHC. The prediction of the helix string model is obtained from the combination of two samples: 
 A (helix-string fragmentation without modification of the particle type content nor the resonance decay)
 and B ( helix-string fragmentation into $\rho$ resonances followed by the resonance decay according to the helix string structure),
 in the proportion 4:1 (the combined sample contains 20\% of the enhanced resonance production). The proportion
 is chosen in such a way that the size of the predicted correlations does not exceed values observed in the LHC data \cite{CMS_be}.

  For completeness, it should be mentioned that a significant improvement in the overall agreement of the MC predictions
 with the inclusive single particle distributions and event shapes measured at LEP has been obtained with the helix-string model
 in an earlier tuning study \cite{tuning}. The study has been repeated with a Pythia8-compatible re-implementation of
  the model~\cite{pystrf}. The results are reported in the Appendix A.

\subsection{Additional experimental input}  
  
  The density of the helix string winding can in principle be measured directly in the
 study of azimuthal correlations of charged hadrons. The experimental data are available
 \cite{ao}  and the comparison with the predictions of the helix string model extended to the decay of short-lived
  resonances is shown in Appendix B. 

   Another experimental input, not discussed so far, may come from the study
 of the polarization of resonance decays. It follows from previous sections that at least
 some of the short-lived resonances should preserve the helix string structure internally and decay
 accordingly, and it should be possible to compare the helix-string model predictions
 with the measurements of the angular distributions in resonance decays.

\subsection{Additional theoretical input}  

   The concept of a helix-shaped confining gluon field offers a fresh insight into certain
 features observed in the data but not sufficiently understood from the theoretical point of view.
 The competition between the helix-string model and the Bose-Einstein interpretation of
 particle correlations is the most striking example, but there are other aspects of the
 problem which merit a theorist's attention. The possibility of the existence of a helix-like ordered gluon
 field emerged from the study of optimal packing of soft gluons at the end of parton cascade, yet it might
 be that some ordered gluon pattern is already present at the level of gluon emission from the leading
 quarks. If this is the case, the properties of the string, transmitted through the correlations of hadrons
 emerging from the interaction, offer a glimpse at the way the charged particle interacts with
 the surrounding field. 
 
   On a more practical level, the study of the helix-string model would profit from further clarification
 of which helix pattern is preferred, or, if the theory does not have a preference, from the implementation of viable options in MC generators so that the model variants can be confronted with the data. In case of the original
  helix-string model proposal~\cite{lund_helixm}, this implies the regularization of the singularity associated with the
 endpoint of the string and the extension of the model on multiparton string configurations.

\section{Conclusions}
   The phenomenological consequences of the replacement of the
 conventional p$_T$ modelling in the Lund string fragmentation 
 with an alternative model based on the helix structure of the QCD string
 are numerous and experimentally verifiable. It is shown how the constraints
 imposed by the helix-shaped gluon field translate into modifications
 of the shape of the inclusive $p_T$ distribution and of
 the momentum difference of hadron pairs. The particle correlations predicted
 by the helix-string model can explain a large part of the enhanced
 production of close pairs of like-sign hadrons. The strength of the model
 consists in its high predictive power, due to a low number
 of free parameters and the reduction of the number
 of degrees of freedom in the modelling.

\appendix
\section*{Appendix A}

  As a cross-check of an earlier tuning study~\cite{tuning}, the predictions of Pythia 8-compatible
 implementation of the helix string model~\cite{pystrf}
 are compared with the data measured by the DELPHI Collaboration~\cite{z0_DELPHI}. The adjustment of model
 parameters is done with help of Rivet~\cite{rivet} and Professor~\cite{professor} packages.

  The study uses as input the inclusive particle spectra and the event shape variables as listed in ~\cite{tuning}.
 The model setup  is identical to Pythia8 default (Tune:ee=3) except for fitted parameters shown in Table I,
 for the standard fragmentation scenario and the helix string fragmentation.
  
  The comparison of the fit results in the table confirms the conclusions reported in~\cite{tuning}, namely that the
overall description of the data improves when the helix string scenario is employed, in particular for the
subset of inclusive particle spectra. The fit provides an estimate of the smearing of the ideal helix shape
(parameter HSF:sigmaHelixRadius, HSF stands for the HelixStringFragmentation class).
 The fitted values of parameters of the helical shape 
 vary with the choice of the input data: the radius (HSF:helixRadius) between 0.4 and 0.54 GeV,
 and the density of the helix winding (HSF:screwiness) between 0.54 and 0.91 rad/GeV.
 Similar spread of values have been obtained in~\cite{tuning} when studying difference between modelling
 using p$_T$-ordered parton shower, and the ARIADNE parton shower.

\begin{table}[h]
\begin{tabular}{|c||c|c|}
\hline
  Tuned parameter &  \multicolumn{2}{|c|}{Input data set} \\
\hline
  Pythia 8   Tuned &  Event shapes \&  &  Inclusive particle \\
 (std.fragm.) &     incl.particle spectra &  spectra only \\
\hline
\hline
 StringPT:sigma & 0.276(1)  &   0.264(3)  \\
 StringZ:aLund  & 0.315(4)  &   0.262(9)  \\
 StringZ:bLund  & 0.689(6)  &   0.60(2)  \\
 TimeShower:alphaSvalue & 0.1418(1) & 0.1452(2) \\
 TimeShower:pTmin &  0.662(8)  & 0.73(3) \\
\hline
 $\chi^2/N_{dof}$ &  3.72 &  3.87 \\
\hline
\hline
  Pythia 8 &  Event shapes \&  &  Inclusive particle \\
 +HELIX~\cite{pystrf} &     incl.particle spectra &  spectra only \\
\hline
\hline
 HSF:screwiness & 0.918(4)  &   0.54(1)  \\
 HSF:helixRadius & 0.405(2)  &   0.53(2)  \\
 HSF:sigmaHelixRadius & 0.063(1)  &   0.07(1)  \\
 StringZ:aLund  & 0.513(3)  &   0.51(6)  \\
 StringZ:bLund  & 0.443(5)  &   0.22(2)  \\
 TimeShower:alphaSvalue & 0.1386(1) & 0.1382(4) \\
 TimeShower:pTmin &  0.767(5)  & 0.74(3) \\
\hline
 $\chi^2/N_{dof}$ &  2.93 &  2.07 \\
\hline
\end{tabular}
\label{tab:tune}
\caption{Results of the tuning study using the DELPHI data~\cite{z0_DELPHI}.}
\end{table}

\begin{figure}[h!]
\begin{center}
\includegraphics[width=0.4\textwidth]{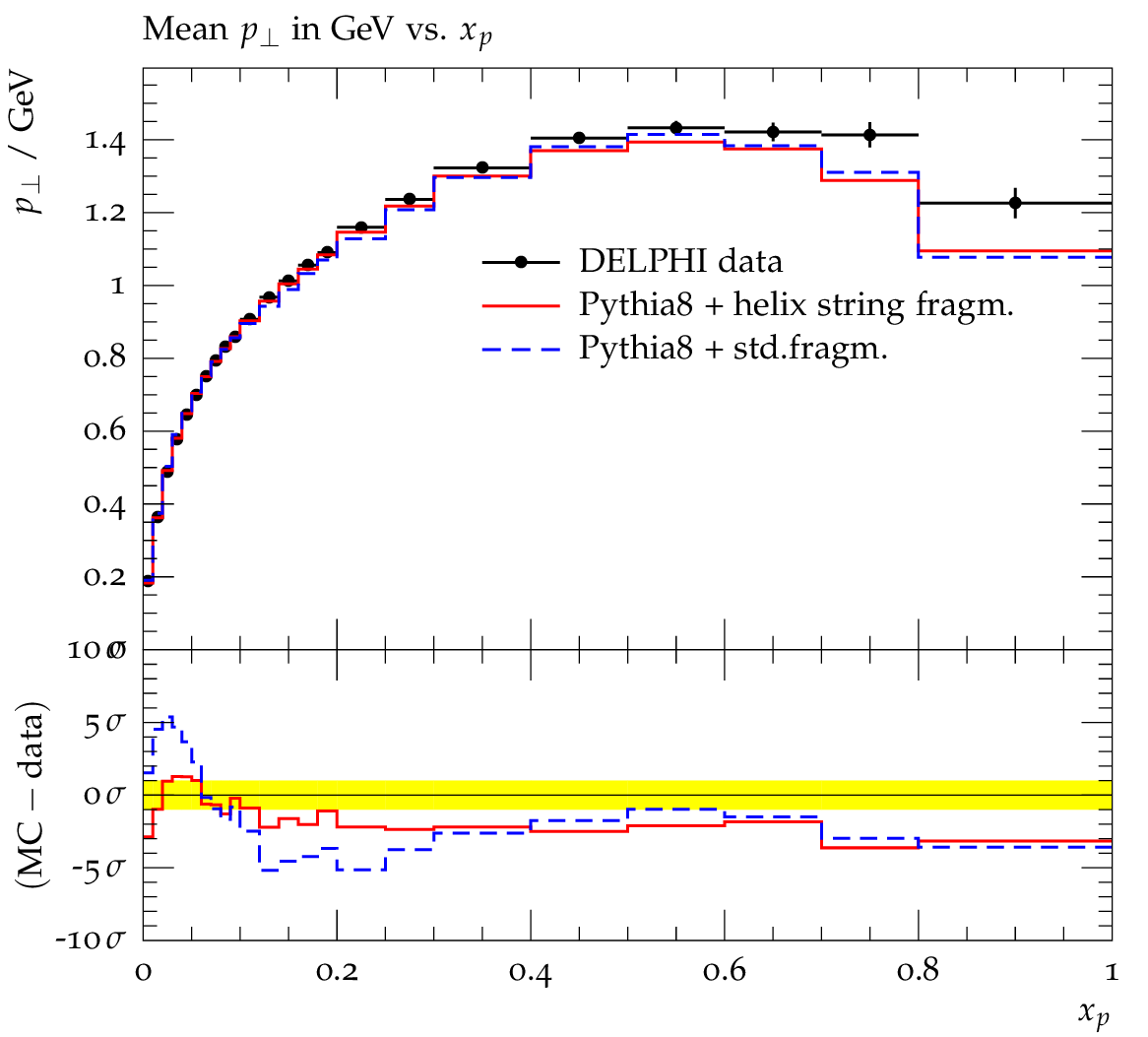}
\caption{ The comparison of the DELPHI data with tuned predictions of Pythia8 using the standard
 fragmentation (dashed blue line) and the helix string fragmentation (full red line).
\label{fig:pt_xp}}
\end{center}
\end{figure}

 Among the inclusive particle spectra, the most significant improvement is obtained in the
 description of the mean transverse momentum measured as the function of the scaled momentum (Fig~\ref{fig:pt_xp}),
 the observable which is best suited to detect the correlations between
 the transverse and absolute momentum of hadrons.
 
\section*{Appendix B}

\begin{table}[b!]
\begin{tabular}{|c||c|}
\hline
  Tuned parameter &  Input data set  \\
\hline
  Pythia8 4C &  ATLAS\_2012\_I1091481  \\
 +HELIX~\cite{pystrf} & inclusive + low-p$_T$ depleted \\
\hline
\hline
 HSF:screwiness & 0.61(2)  \\
 HSF:helixRadius & 0.460(5)  \\
 MultipartonInteractions:expPow &  3.7(1) \\
 MultipartonInteractions:pT0Ref &  2.09(3) \\
 SpaceShower:pT0Ref &  1.81(4) \\
 BeamRemnants:reconnectRange & 0.[F] \\
\hline
 $\chi^2/N_{dof}$ & 581/443 = 1.3 \\
\hline
\end{tabular}
\label{tab:tune_atlas}
\caption{Results of the tuning study using the ATLAS data~\cite{ao} collected at $\sqrt{s}$=7 TeV.}
\end{table}

  Recently, the ATLAS Collaboration published the measurement of the azimuthal ordering
 of hadrons inspired by the search of the signature of the helix-like shaped QCD field~\cite{ao}.
 The spectral analysis of correlations between the azimuthal opening angle and the longitudinal separation
 of hadrons provides a possibility to measure the parameter describing the helix winding density in the model. 
 The data show presence of correlations in the region of interest (known from the previous study of
 hadronic Z$^0$ data). The interpretation of the observed effect is complicated by the fact that
 the discrepancies between the data and the predictions of conventional models cannot be attributed
 to the modelling of hadronisation alone but require re-adjustment in the jet sector (sensitive
 to the amount of multiple parton interactions/MPI, ISR, and colour reconnection/CR). 
 
\begin{figure}[b!]
\begin{center}
\includegraphics[width=0.45\textwidth]{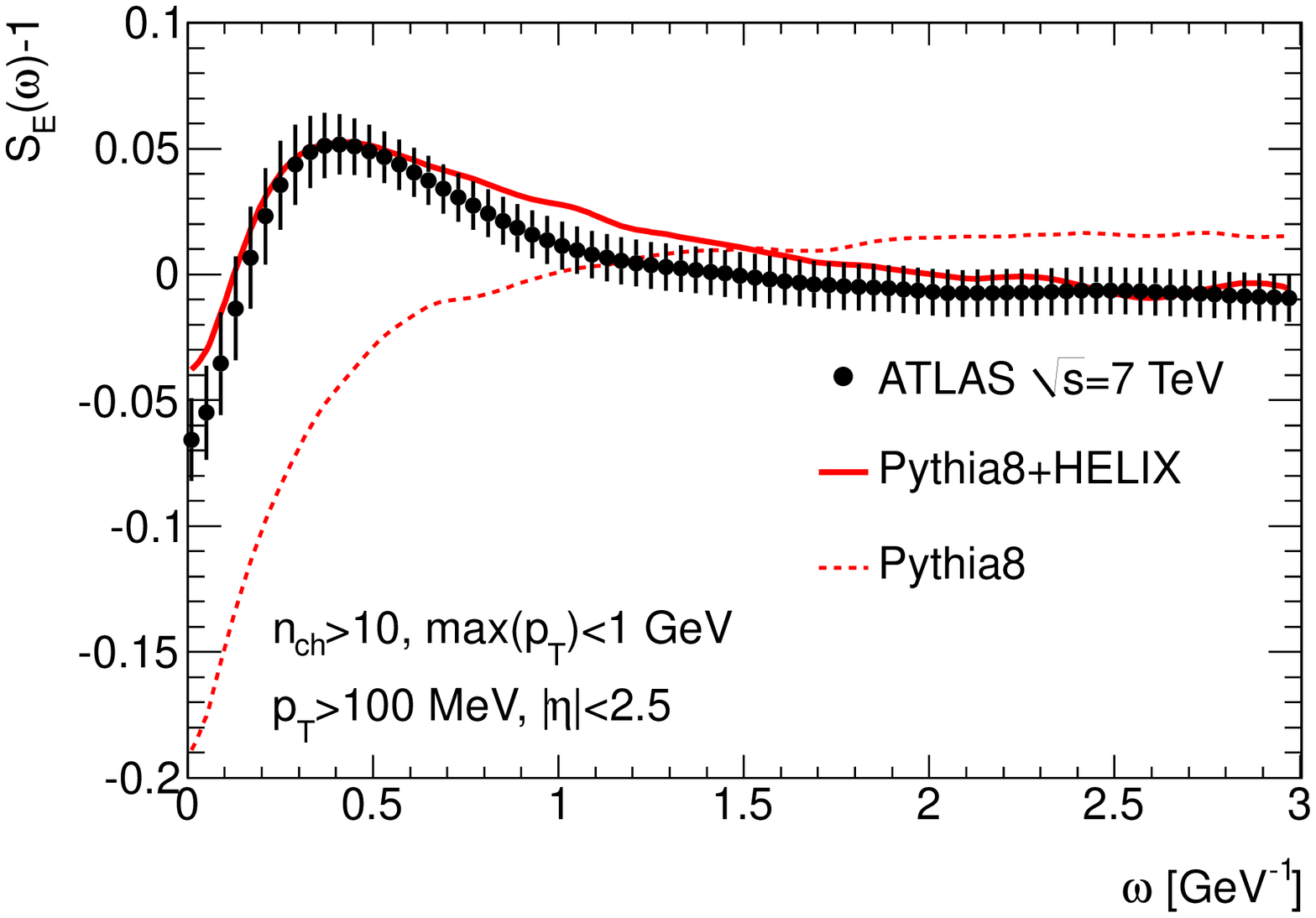}
\includegraphics[width=0.45\textwidth]{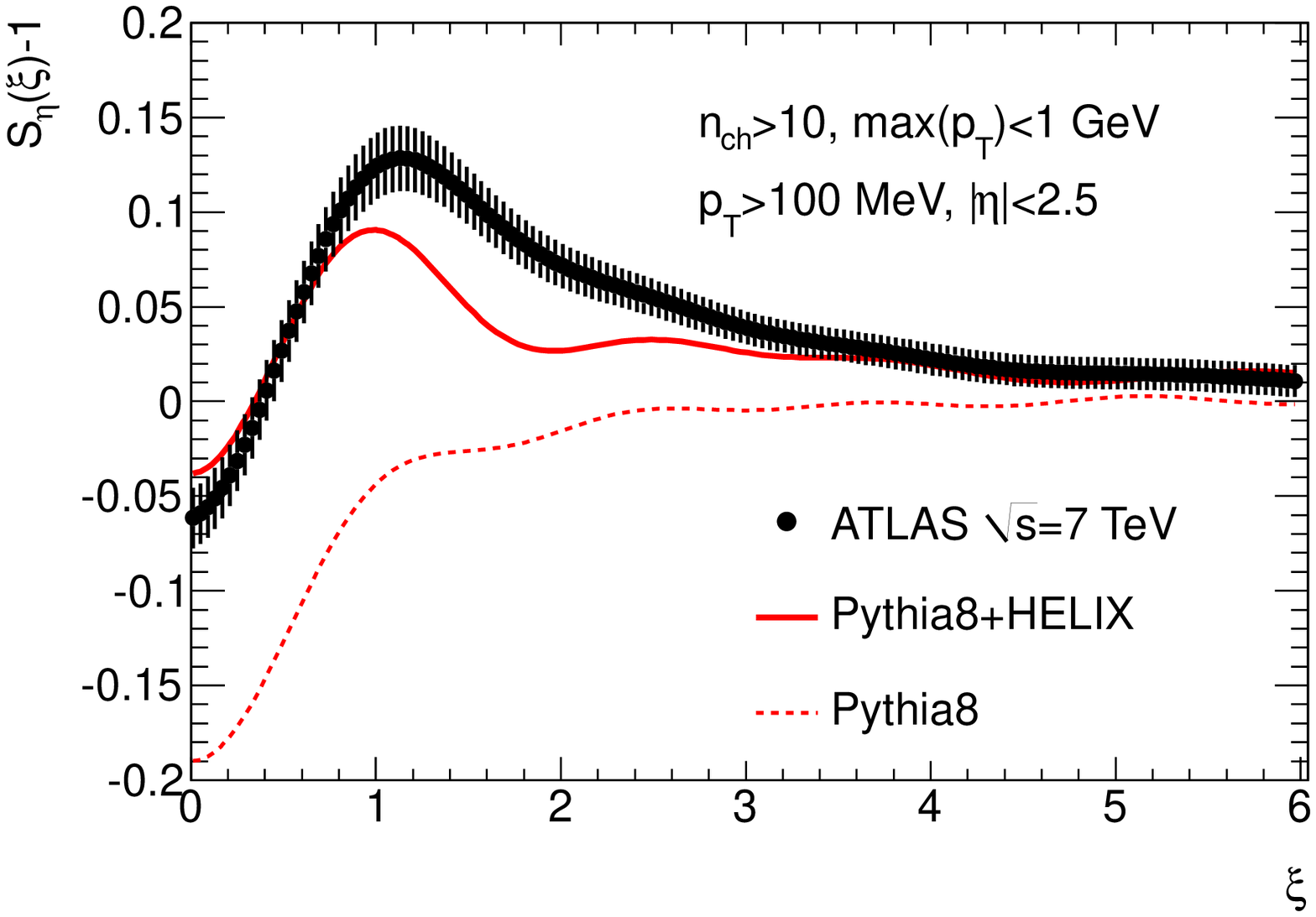}
\caption{ The comparison of the ATLAS data with tuned predictions of Pythia8 using the
 helix string model extended to the decay of short-lived resonances (full line). To illustrate the effect
 of the helix string structure, the predictions of the model with the standard string fragmentation and resonance decay
 are indicated (dashed line), for the same parameter setup. Low-p$_T$ enhanced event selection (restricted by the
 cut on the max(p$_T)<$ 1 GeV).
\label{fig:ao}}
\end{center}
\end{figure}

   According to the arguments developed in this paper, the helix string model should be extended to the decay
 of short-lived resonances, which means a significant increase of helix-string induced correlations in the model
 prediction. It is interesting to see how well these predictions compare with the measurement. 
 
   The study is done in the following way: the Pythia~8 based implementation of the helix string model
 is used with the default hadron mixture. The $\rho$ and K$^*$ resonances decay according to their
 'internal' helix string structure. The parameters describing parton shower,fragmentation function and
 the smearing of the helix radius are fixed to values obtained in the fit of Z$^0$ data (Appendix A). 
 A selection of parameters describing MPI, ISR and CR (see Table II) is retuned in order to achieve
 a good agreement between the data and the model prediction in the inclusive event selection and
 in the low-p$_T$ depleted region, less sensitive to hadronisation effects. The main parameters of
 the helix structure are included in the fit and their optimized values are within the range obtained
 from the comparison of the model with Z$^0$ data. Since the data indicate a very small amount of colour reconnection,  
 the reconnection probability is set to 0 to stabilize the minimisation procedure.   

 The predictions of the retuned model for the region most sensitive to the hadronisation effects
 (low-p$_T$ enhanced region) are shown in Fig~\ref{fig:ao}. A reasonable agreement between the data
 and the model is seen in the S$_E$ oservable. The model somewhat underestimates the size of correlations
 in the  S$_\eta$ power spectrum, and the predicted distribution shows a wavy structure. This is an indication
 that some of spectral components are missing in the simulation and it provides a hint for further improvement
 of the modelling (the origin of the discrepancy in the $S_\eta$ may be related to a poorly adjusted 
 distribution of the invariant mass of hadronic systems).   

  For comparison, the predictions of Pythia~8 with the same parameter setup but using the standard fragmentation and decay
 algorithm are shown together with the predictions of the helix string model. The difference serves as an
 illustration of the effect the helix string model has on the size of the azimuthal ordering signal.

\bibliography{draft}

\end{document}